\documentclass[aps,preprint,showpacs,superscriptaddress,groupedaddress]{revtex4}  
\usepackage{appendix}
\usepackage{graphicx}  
\usepackage{dcolumn}   
\usepackage{bm}        
\usepackage{amssymb}   
\usepackage{amsmath}
\usepackage{array}
\usepackage{mathalfa}
\usepackage{epstopdf}
\usepackage{mathtools}
\usepackage[T1]{fontenc}  
\usepackage{textcomp}     
\usepackage{mathtools,slashed}
\usepackage{graphics}
\usepackage{epsfig}
\usepackage{psfrag}
\usepackage{color}
\usepackage[utf8]{inputenc}
\usepackage{scrextend}
\usepackage{setspace}
\DeclareMathAlphabet{\pazocal}{OMS}{zplm}{m}{n}
\usepackage{amsfonts}
\usepackage{amsthm}
\usepackage{float}
\usepackage{textcomp}
\usepackage[section]{placeins}
\usepackage[caption=false]{subfig}
\usepackage{subfig}
\usepackage{tabularx}

\begin{document}
\title{Proton Decay in Supersymmetric $\bm{SU(4)_c\times SU(2)_L \times SU(2)_R}$}
\date{\today}
\author{George Lazarides}
\email[E-mail: ]{lazaride@eng.auth.gr}
\affiliation{School of Electrical and Computer Engineering, Faculty of Engineering,
Aristotle University of Thessaloniki, Thessaloniki 54124, Greece}
\author{Mansoor Ur Rehman}
\email[E-mail: ]{mansoor@qau.edu.pk}
\affiliation{Department of Physics, Quaid-i-Azam University, Islamabad 45320, Pakistan}
\author{Qaisar Shafi}
\email[E-mail: ]{shafi@bartol.udel.edu}
\affiliation{Bartol Research Institute, Department of Physics and Astronomy, University of Delaware, Newark, DE 19716, USA}

\begin{abstract}
We discuss proton decay in a recently proposed model of supersymmetric hybrid inflation based on the gauge symmetry $SU(4)_c \times SU(2)_L \times SU(2)_R$. A $U(1)\, R$ symmetry plays an essential role in realizing inflation as well as in eliminating some undesirable baryon number violating operators. Proton decay is primarily mediated by a variety of color triplets from chiral superfields, and it lies in the observable range for a range of intermediate scale masses for the triplets. The decay modes include $p \rightarrow e^{+}(\mu^+) + \pi^0$, $p \rightarrow \overline{\nu} + \pi^{+}$, $p \rightarrow K^0 + e^+(\mu^{+})$, and $p \rightarrow K^+ + \overline{\nu}$, with a lifetime estimate of order $10^{34}-10^{36}$ yrs and  accessible at Hyper-Kamiokande and future upgrades. The unification at the Grand Unified Theory (GUT) scale $M_{\rm GUT}$ ($\sim 10^{16}$~GeV) of the Minimal Supersymmetric Standard Model (MSSM) gauge couplings is briefly discussed.
\end{abstract}

\maketitle
\section{\label{Intro}Introduction}

In a recent paper \cite{Lazarides:2020zof} we proposed a realistic supersymmetric hybrid inflation scenario \cite{Dvali:1994ms,Copeland:1994vg} specifically tailored for the gauge symmetry $SU(4)_c \times SU(2)_L \times SU(2)_R$ ($G_{\text{4-2-2}}$) \cite{Pati:1973uk,Pati:1973rp,Pati:1974yy}. The model employs shifted hybrid inflation \cite{Jeannerot:2000sv,Jeannerot:2001xd} during which the $G_{\text{4-2-2}}$ symmetry is broken and the doubly charged monopoles \cite{magg} are inflated away. The model is fully compatible with the Planck data \cite{Akrami:2018odb} and, for a wide choice of parameters, it predicts observable gravity waves generated during the inflationary epoch. The $G_{\text{4-2-2}}$ symmetry breaking scale is estimated to be of order $M_{\rm GUT}$ ($\sim 10^{16}$~GeV).

Motivated by the above development in this follow-up paper we explore the important issue of proton decay in these supersymmetric $G_{\text{4-2-2}}$ models. It is well known that such models do not contain any superheavy gauge bosons that can mediate proton decay. However, proton decay in our 
$G_{\text{4-2-2}}$ model can arise from the exchange of a variety of color triplets present in the various chiral superfields. With intermediate scale masses of varying range that we estimate for these states, the proton decay rate is found to be accessible in the next generation experiments such as JUNO \cite{An:2015jdp}, DUNE \cite{Acciarri:2015uup}, and Hyper-Kamiokande \cite{Abe:2018uyc}.

The layout of the paper is as follows. In Sec.~II we describe the superpotential and the field content of our model. A $U(1)\, R$ symmetry, which is required to realize hybrid inflation, is also shown to play an important role in eliminating some undesirable baryon number violating operators in Sec.~III. In addition, we discuss the possibility of observable proton decay with the intermediate mass scale color triplets and a successful realization of MSSM gauge coupling unification with additional bi-doublets. Our conclusions are summarized in Sec.~IV.
\section{\label{Mod}Supersymmetric $\bm{SU(4)_c\times SU(2)_L \times SU(2)_R}$ model}
The MSSM matter superfields including right-handed neutrinos ($\nu^c$) are contained in $F$ and $F^c$ belonging to the following representations:
\begin{equation}
F_i=(4, 2,1)\equiv
  \left( {\begin{array}{cccc}
   u_{ir} &  u_{ig}  &  u_{ib}  & \nu_{il} \\
   d_{ir} &  d_{ig}  &  d_{ib}  & e_{il} \\
  \end{array} } \right),\ \
   F^c_i=(\overline 4, 1, 2)\equiv 
  \left( {\begin{array}{cccc}
   u^c_{ir} &  u^c_{ig}  &  u^c_{ib}  & \nu^c_{il} \\
   d^c_{ir} &  d^c_{ig}  &  d^c_{ib}  & e^c_{il} \\
    \end{array} } \right),\\
\end{equation}
where $i=1$, 2, 3 is the generation index, and the subscripts $r,\ g,\ b,\ l$ represent the four colors in the model, namely red, green, blue, and lilac. 
It is sufficient to consider a right-isospin-doublet four-colored GUT Higgs superfield $H^c$ and its conjugate superfield $\overline{H^c}$ with the following representations: 
 \begin{equation}\label{vev}
H^c=(\overline4, 1, 2)\equiv 
  \left( {\begin{array}{cccc}
   u^c_{Hr} &  u^c_{Hg}  &  u^c_{Hb}  & \nu^c_{Hl}\\
   d^c_{Hr} &  d^c_{Hg}  &  d^c_{Hb}  & e^c_{Hl} \\
  \end{array} } \right),\ \
  \overline {H^c}=(4, 1, 2)\equiv 
  \left( {\begin{array}{cccc}
   \overline {u^c_{Hr}} &  \overline{u^c_{Hg}}  & \overline{u^c_{Hb}}  & \overline{ \nu^c_{Hl}}\\\
   \overline {d^c_{Hr}} & \overline {d^c_{Hg}}  & \overline{d^c_{Hb}}  &\overline {e^c_{Hl}} \\
    \end{array} } \right),\\
\end{equation}
in order to achieve the breaking of the $G_{\text{4-2-2}}$ gauge symmetry to the Standard Model (SM) gauge symmetry $G_{\rm SM}=SU(3)_c\times SU(2)_L\times U(1)_Y$. These fields acquiring nonzero vacuum expectation values (vevs) along the right-handed sneutrino directions, that is $|\langle \nu^c_{Hl}\rangle|=|\langle \overline{ \nu^c_{Hl}}\  \rangle |=v\neq0$, with $v$ around the GUT scale ($\sim 2\times10^{16}$~GeV). The electroweak  breaking is triggered by the electroweak Higgs doublets $h_u$ and $h_d$ residing in the bi-doublet Higgs superfield $h$ represented by 
\begin{equation}
h=(1, 2, 2)\equiv (h_u \ \ h_d)= \left( {\begin{array}{cc}
    h^+_u &   h^0_d   \\
  h^0_u &     h^-_d   \\
  \end{array} } \right).\\
\end{equation}
Such doublets can remain light as a result of appropriate discrete symmetries \cite{lightdoublets}. 
Due to an $R$ symmetry the color triplet pair $d^c_H$ and $\overline{d^c_H}$ remains massless. An economical choice to remedy this problem is the introduction of a sextet superfield $G = (6, 1, 1)$ with SM components $g = (3, 1, -1/3)$ and $g^c = (\overline 3, 1, 1/3)$. This can provide superheavy masses to the color triplets $d^c_H$ and $\overline{d^c_H}$ by mixing them with $g$ and $g^c$ 
\cite{King:1997ia}. Finally, to realize inflation within the supersymmetric hybrid framework a gauge singlet chiral superfield $S = (1, 1, 1)$ is introduced whose scalar component plays the role of the inflaton. The various superfields with their representation, transformation under $G_\text{4-2-2}$, decomposition under $G_{\rm SM}$, and respective charge $q(R)$ are shown in Table~\ref{assign1}. 

\begin{table}
\caption{\label{assign1} Field content together with their decomposition under the SM and $R$ charge.}
\resizebox{1.0\textwidth}{!}{
\begin{ruledtabular}
\begin{tabular}{| >{\centering\arraybackslash}m{1in}| | >{\centering\arraybackslash}m{1in} |>{\centering\arraybackslash}m{2in} | >{\centering\arraybackslash}m{1in} |  }
Superfields&$4_c\times 2_L \times2_R$ & $3_c\times 2_L \times1_Y$     &$q(R)$  \\
\hline 
\hline  
 \vspace{0.2cm}
$F_i$& $({4,\ 2,\ 1})$  & $Q_{ia} ({ 3,\ 2},\ \ \ 1/6)$&1 \\
&  & $L_i ({ 1,\ 2},\ -1/2)$&  \\
\hline
 \vspace{0.2cm}
$F^c_i$& $({\overline{4},\ 1,\ 2})$ & $u^c_{ia}({ \overline{3},\ 1},\ -2/3)$&1  \\
&   & $d^c_{ia}({ \overline{3},\ 1},\ \ \ 1/3)$&  \\
&  &$\nu^c_i \ ({ 1,\ 1},\ \ 0)$&\\
&   & $e^c_i \ ({ 1,\ 1},\ \ 1)$&\\
\hline
\hline
 \vspace{0.2cm}
 $H^c$& $ ({\overline{4},\ 1,\ 2})$ & $u^c_{Ha} ({ \overline{3},\ 1},\ -2/3)$&0\\
&  & $d^c_{Ha}({ \overline{3},\ 1},\ \ \ 1/3)$&\\
&  &$\nu^c_H\  ({ 1,\ 1},\ \  0)$&\\
&  & $e^c_H \ ({ 1,\ 1},\ 1)$&\\
\hline
 \vspace{0.2cm}
$\overline {H^c}$& $ ({4,\ 1,\ 2})$ & $\overline {u^c_{Ha}} ({ 3,\ 1},\ \ \ 2/3)$&0  \\
&  & $\overline {d^c_{Ha}} ({3,\ 1},\ \ \ -1/3)$& \\
& & $\overline {\nu^c_H}\  ({ 1,\ 1},\ \  0)$& \\
& &$\overline {e^c_H}\  ({ 1,\ 1},\  -1)$& \\
\hline
 \vspace{0.2cm}
$ S$& $ ({1,\ 1,\ 1})$ & $ S({ 1,\ 1},\ \ \ 0)$&2 \\
\hline
 \vspace{0.2cm}
$ G$& $ ({6,\ 1,\ 1})$ & $g_{a} ({3,\ 1},\ -1/3)  $&2 \\
$ $&  &$ g^c_{a} ({\overline{3},\ 1},\  \  \ 1/3)$& \\
\hline
 \vspace{0.2cm}
 $h$&$({1,\ 2,\ 2})$ & $h_u\  ({1,\ 2},\ \ \ 1/2)$&0 \\
&  &$h_d\ ({ 1,\ 2},\ -1/2)$&  \\
\end{tabular}
\end{ruledtabular}
}
\end{table}

It can be noted from the Table~\ref{assign1} that the MSSM matter superfields $F,\,F^c$ carry one unit of $R$ charge, while the MSSM Higgs doublets in $h$ are neutral under the $R$ symmetry. This reflects the fact that the matter-parity $\mathbb Z_2^{mp}$, which is usually invoked to forbid rapid proton decay operators at the renormalizable level, is contained in $U(1)_{R}$ as a subgroup. The superpotential $W$ is invariant under $\mathbb Z_2^{mp}$ and this symmetry remains unbroken. Therefore, no domain wall problem appears here and consequently the lightest supersymmetric particle (LSP) becomes a plausible dark matter candidate. It is interesting to note that a $Z_4$ subgroup of $U(1)_R$ symmetry \cite{Lee:2010gv}, consistent with the $R$ charge assignment displayed in the Table~\ref{assign1}, plays a key role in constraining the possible superpotential terms.

The superpotential employed in Ref.~\cite{Lazarides:2020zof} for the shifted $\mu$-hybrid inflation with $G_{\text{4-2-2}} \times U(1)_R$ symmetry is given by
\begin{align}\label{SP1}
W & = \kappa S (\overline{H^c}H^c - M^2)+\lambda Sh^2 - S \beta \frac {(\overline{H^c} H^c)^2}{\Lambda^2} \nonumber \\ 
& + a\,GH^cH^c + b\,G\overline{H^c}\ \overline{H^c} + \lambda_{ij}F^c_iF_j h  \nonumber \\
& + \left( \gamma_{1}^{ij} F^c_i F^c_j + \gamma_{2}^{ij} F_i F_j \right) \frac{H^c H^c}{\Lambda} + \left( \overline{\gamma}_{1}^{ij} F^c_i F^c_j + \overline{\gamma}_{2}^{ij} F_i F_j \right) \frac{\overline{H^c} \, \overline {H^c}}{\Lambda},
\end{align}
where $\kappa,\ \lambda,\ \beta,\ a,\ b,\ \lambda^{ij}_{1,2},\  \gamma^{ij}_{1,2},\ 
\text{ and } \overline{\gamma}^{ij}_{1,2}$ are real and positive dimensionless couplings and $M$ is a superheavy mass parameter. The superheavy scale $\Lambda$ is assumed to lie in the range $10^{16}$~GeV$\lesssim \Lambda \lesssim m_P$, where $m_P \simeq 2.4\times 10^{18}$~GeV is the reduced Planck mass.  The first-line terms in the superpotential $W$ in Eq.~(\ref{SP1}) are relevant for the shifted $\mu$-hybrid inflation and the resolution of the monopole problem, as discussed in Ref.~\cite{Lazarides:2020zof}. In addition, the coupling $\lambda S h_u h_d$ yields the MSSM $\mu$ term once the scalar component of the superfield $S$ acquires a nonzero vev proportional to the gravitino mass $m_{3/2}$ with $\mu = -\lambda m_{3/2}/\kappa$ \cite{Dvali:1997uq}. The achievement of low reheat temperatures $T_r\gtrsim 10^5$~GeV, the possibly observable gravity waves with tensor-to-scalar ratio $r\lesssim 10^{-4}-10^{-3}$, and the gravitino dark matter with inflationary predictions consistent with the latest Planck data are the attractive features of this inflationary model as discussed in detail in Ref.~\cite{Lazarides:2020zof}. For earlier  work on the $\mu$-hybrid inflation model see Refs.~\cite{Okada:2015vka} and \cite{Rehman:2017gkm}.

The first two terms in the second line of Eq.~(\ref{SP1}), which include the sextet superfield $G $, serve to provide superheavy masses to $d^c_H$ and $\overline{d^c_H}$ as discussed above. 
The Yukawa interactions of the matter superfields $F,\,F^c$ are represented by the $\lambda_{ij}$-couplings. The neutrino ($\nu$) and right-handed neutrino ($\nu^c$) couplings from the $\lambda_{ij}$- and $\gamma^{ij}_1$-terms explain the tiny neutrino masses via the see-saw mechanism. The $\gamma^{ij}$- and $\overline{\gamma^{ij}}$-couplings in the third line of $W$ play an important role in generating possibly observable proton decay as discussed in the next section in detail.

\section{\label{PD}Proton decay in R-symmetric $\bm{SU(4)_c\times SU(2)_L \times SU(2)_R}$ model}
The fact that the gauge bosons in the $G_{\text{4-2-2}}$ model do not mediate proton decay seems to support the observed stability of proton. We therefore only discuss proton decay mediated via the color triplets present in the chiral superfields $F,\,F^c \supset d,\,d^c$, $G = g + g^c$, and 
$H^c,\,\overline{H^c} \supset d^c_H,\, \overline{d^c_H}$. 
This mediation can effectively generate four-Fermi proton decay operators with chirality type 
LLLL, RRRR, or LLRR. As discussed below, the $R$ symmetry does not allow four-Fermi operators of the type LLLL and RRRR, whereas observable proton decay is only mediated through the color triplets $d^c_H,\, \overline{d^c_H}$ with four-Fermi operators of LLRR chirality.

\subsection{R-symmetry Breaking Proton Decay Modes}

The dimension-four $L$- and $B$-violating operators may appear at the nonrenormalizable level in the superpotential as
\begin{equation}
\frac{FFF^cH^c}{\Lambda} \supset  \frac{v}{\Lambda}(LLe^c + QLd^c), \quad F^cF^cF^cH^c \supset  \frac{v}{\Lambda} u^c d^c d^c,
\end{equation}
which can lead to fast proton decay via the effective operator $(v/\Lambda)^2QL(u^c)^{\dagger}(d^c)^{\dagger}$, suppressed by the color-triplet $d^c$-squark mass. However, these operators are not allowed by the $R$~symmetry defined in Table~I. Similarly, the dimension-five $L$- and $B$-violating operators arising from the following  nonrenormalizable gauge invariant terms in the superpotential 
\begin{equation}
\frac{FFFF}{\Lambda} \supset \frac{Q\,Q\,Q\,L}{\Lambda}, \quad \frac{F^cF^cF^cF^c}{\Lambda}  \supset \frac{(u^c u^c d^c e^c + u^c d^c d^c \nu^c)}{\Lambda}
\end{equation}
are forbidden by the $R$~symmetry. The gauge invariant renormalizable interactions $FFG \supset QQg+LQ g^c$, or $F^c F^c G \supset u^c d^c g^c + u^c e^c g + d^c \nu^c g$ with the sextet $G$ can also mediate dimension-five fast proton decay via the chirality flipping propagator with a $GG \supset g^c g$  mass insertion \cite{King:1997ia,Shafi:1998yy}. Again, the $R$~symmetry does not allow these terms in the superpotential which could otherwise generate LLLL and RRRR four-Fermi operators.
\begin{figure}[t]
\begin{tabularx}{0.8\textwidth}{@{}cXX@{}}
\begin{tabular}{cc}
\hspace{1.3cm}\subfloat[]{\includegraphics[scale=0.6]{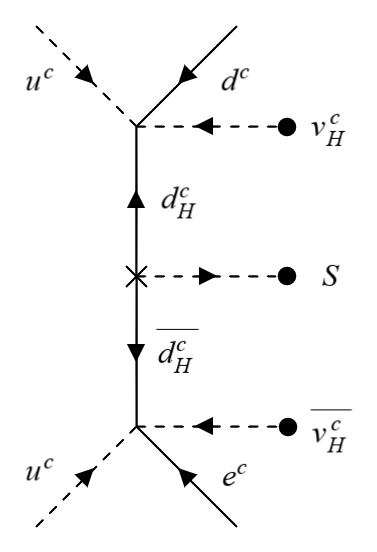}} \hspace{1.0cm}&\subfloat[]{\includegraphics[scale=0.6]{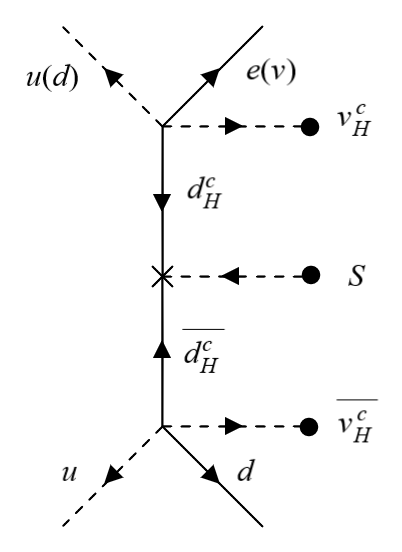}}  
\end{tabular}
\end{tabularx}
\caption{\label{pd51} Dimension-five proton decay diagrams. Dashed lines represent bosons, solid lines represent fermions, and dots represent the vevs. These are graphs between the $B$-violating superpotential coupling $F^cF^cH^cH^c$ ($FF\overline{H^c}\ \overline{H^c}$) and the $B$-conserving superpotential coupling $F^cF^c\overline{H^c}\ \overline{H^c}$ ($FFH^cH^c$). The fermionic or bosonic character of the external lines in each vertex can be interchanged independently.} 
\end{figure}

The breaking of the $R$ symmetry in the hidden sector can also assist proton decay via the soft supersymmetry breaking terms, although the corresponding decay rates are generally expected to be suppressed. As an example, consider the following $R$-symmetric nonrenormalizable terms
\begin{equation} \label{gij}
W \supset \left( \gamma_{1}^{ij} F^c_i F^c_j + \gamma_{2}^{ij} F_i F_j \right) \frac{H^c H^c}{\Lambda} + \left( \overline{\gamma}_{1}^{ij} F^c_i F^c_j + \overline{\gamma}_{2}^{ij} F_i F_j \right) \frac{\overline{H^c} \, \overline {H^c}}{\Lambda}.
\end{equation} 
These interactions yield effective dimension-five proton decay operators via a chirality flipping propagator involving a mass insertion $\kappa \langle S \rangle \overline{H^c}H^c = -m_{3/2} \overline{H^c}H^c $, as shown in Fig.~\ref{pd51}. Here, the solid lines refer to fermions, the dashed lines to bosons, and the dotes represent the vevs. The $S$ field acquires a nonzero vev due to the 
violation of the $R$ symmetry by the soft supersymmetry breaking terms \cite{Dvali:1997uq}. In Fig.~\ref{pd51} and thereafter we use the same notation for the chiral superfields and their scalar and fermionic components.

To provide an order of magnitude estimate for the proton decay rate we assume all dimensionless coupling constants in Eq.~(\ref{gij}) to be of the same order. Note that only the $\overline{\gamma}_{1}^{ij}\equiv\gamma^{ij}$ coupling is actually related to the right-handed neutrino Majorana mass matrix $\gamma_{ij}(v^2/\Lambda)$ with eigenvalues $M_i = \gamma_i (v^2/\Lambda)$. The distinguishing feature of linking proton decay to neutrino masses via the right-handed neutrino Majorana mass terms is highlighted in Refs.~\cite{Babu:1997js,Babu:1998wi,Pati:2002ig}. Connecting external squark and/or slepton lines, in each of these dimension-five diagrams with a Higgsino or gaugino line one can generate one-loop (box) diagrams corresponding to LLLL and RRRR type four-Fermi proton decay operators.
Other possible diagrams with an external $\nu^c$ line are not allowed kinematically, whereas the amplitude for diagrams with internal $\nu_H^c$ lines is suppressed as compared to the diagrams shown in Fig.~\ref{pd51}. Assuming all $\gamma_{1,2}^{ij}$'s and $\overline{\gamma}_{1,2}^{ij}$'s  to be of order $\gamma_i$, the amplitude of the box diagrams corresponding to the dimension-five diagrams in Fig.~\ref{pd51} contains the suppression factor $(m_{3/2}/m_{d^c_H} m_{\overline{d^c_H}})^2 (M_i/v)^2 $, where $m_{d^c_H}= a\,v$, $m_{\overline{d^c_H}}= b\,v$ and $\mu=(-\lambda/\kappa) m_{3/2} \sim m_{3/2}$ is assumed. Due to color antisymmetry of the relevant dimension-five operators the dominant proton decay mode is $p \rightarrow \overline{\nu} K^+$ with a corresponding lifetime bound  $\tau_{p \rightarrow \overline{\nu} K^+} \gtrsim 6.6 \times  10^{33}$ yrs \cite{Abe:2014mwa,Takhistov:2016eqm} from the Super-Kamiokande experiment. For a given value of $M_i/v$ and the MSSM parameter $\tan\beta$ and assuming that the box diagrams with a Higgsino exchange dominate, this translates into a lower bound on the masses of the color triplets $d^c_H,\overline{d^c_H}$:
\begin{equation} \label{bd5}
\sqrt{m_{d^c_H} m_{\overline{d^c_H}}} \gtrsim 1.6 \times  10^{8} \left(\frac{m_{3/2}/\sqrt{\sin 2\beta}}{10^3 \text{ GeV}} \right)^{1/2} \left( \frac{M_i}{v} \right)^{1/2}~\text{ GeV}.
\end{equation}
For typical values of $m_{3/2} \sim $ TeV, $\tan \beta \sim 10$, and $v = 2 \times 10^{16}$ GeV, we obtain the largest lower bound $\sqrt{m_{d^c_H} m_{\overline{d^c_H}}} \gtrsim 1.7 \times 10^{7}$ GeV or $\sqrt{ab} \gtrsim 10^{-9}$, corresponding to the heaviest right-handed neutrino mass $M_i \sim 10^{14}$ GeV. Assuming natural values for the couplings $a,\,b\sim 1$, this decay rate is highly suppressed. 

\begin{figure}[t]
\begin{center}
\begin{tabularx}{0.8\textwidth}{@{}cXX@{}}
\begin{tabular}{cc}
\hspace{1.3cm}\subfloat[]{\includegraphics[scale=0.6]{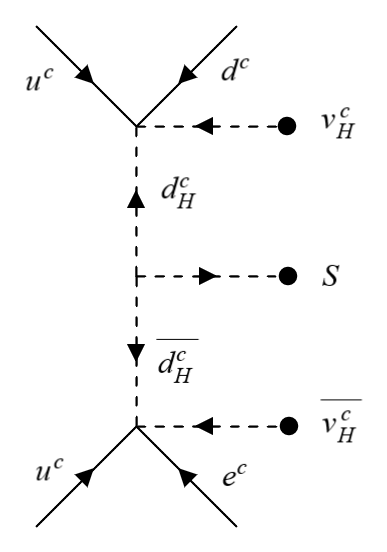}} \hspace{1.0cm}&\subfloat[]{\includegraphics[scale=0.6]{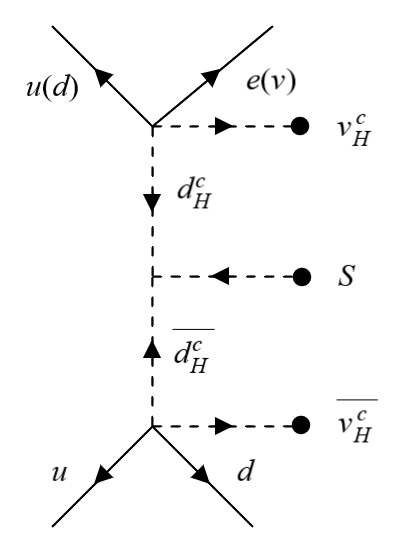}} 
\end{tabular}
\end{tabularx}
\end{center} 
\caption{\label{pd6cf} Dimension-six proton decay diagrams with chirality flipping mediation. Dashed lines represent bosons, solid lines represent fermions, and dots represent the vevs. These are graphs between the $B$-violating superpotential coupling $F^cF^cH^cH^c$ ($FF\overline{H^c}\ \overline{H^c}$) and the $B$-conserving superpotential coupling $F^cF^c\overline{H^c}\ \overline{H^c}$ ($FFH^cH^c$).} 
\end{figure}

For the sake of completeness, we briefly discuss  proton decay via the dimension-six operators of type RRRR and LLLL represented by the diagrams shown in Fig.~\ref{pd6cf}. The scalar cubic coupling involving the three relevant fields $S, \, \overline {H^c},\, H^c$ is provided by the soft supersymmetry breaking trilinear coupling $\kappa A m_{3/2} \, S \, \overline {H^c} H^c+{\rm h.c}$, where $A$ is a dimensional constant of order unity. The current limit on proton partial lifetime, $\tau (p\rightarrow \pi^0 l^+) > 1.6 \times 10^{34}$ yrs \cite{Miura:2016krn}, then yields the following lower bound on the masses of the color triplets $d^c_H,\overline{d^c_H}$:
\begin{equation} \label{bd6}
\sqrt{m_{d^c_H} m_{\overline{d^c_H}}} \gtrsim 1.6 \times 10^{9} \left( \frac{M_i}{v} \right)^{1/2} \left( \frac{m_{3/2}}{10^3 \text{ GeV}} \right)^{1/2}
~\text{ GeV}.
\end{equation}
For $m_{3/2} \sim $ TeV and $v = 2 \times 10^{16}$ GeV, we obtain the largest lower bound $\sqrt{m_{d^c_H} m_{\overline{d^c_H}}} \gtrsim 10^{8}$~GeV or $\sqrt{ab} \gtrsim 5.6 \times 10^{-9}$, corresponding to the heaviest right-handed neutrino mass $M_i \sim 10^{14}$ GeV. This value is roughly comparable to the value obtained above from the dimension-five proton decay (i.e., $\sqrt{ab} \gtrsim 10^{-9}$).

\subsection{R-symmetric Observable Proton Decay Modes}

We now discuss dimension-five and dimension-six proton decay operators of type LLRR which are generated from the interference of the interactions in $\int d^2\theta\, W$ with their Hermitian conjugates. After integrating out the heavy color triplets, the effective operators obtained fall into the category of the following four-Fermi operators:
\begin{equation}
FF (F^c)^{\dagger}(F^c)^{\dagger} \supset QQ (u^c)^{\dagger} (e^c)^{\dagger} + QL(u^c)^{\dagger}(d^c)^{\dagger}. \label{LLRR}
\end{equation}
The $R$ symmetry is automatically respected by these operators and the proton decay rates can be predicted in the observable range without the $R$-symmetry breaking suppression factors. 

\begin{figure}[th]
\begin{tabularx}{0.8\textwidth}{@{}cXX@{}}
\begin{tabular}{ccc}
\subfloat[]{\includegraphics[scale=0.32]{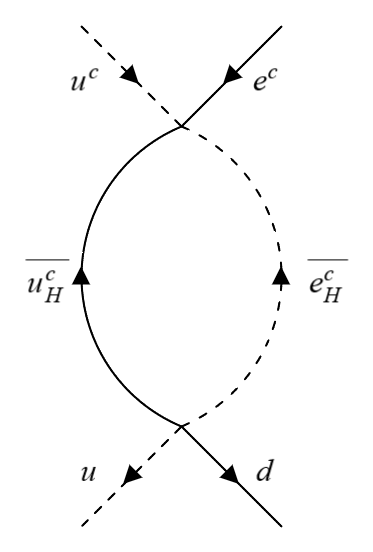}} \hspace{0.5cm}&\subfloat[]{\includegraphics[scale=0.32]{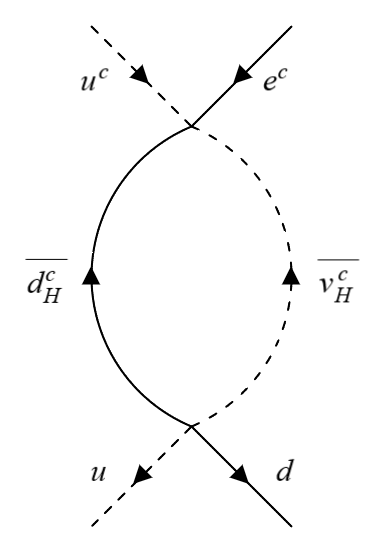}} \hspace{0.5cm} & \subfloat[]{\includegraphics[scale=0.52]{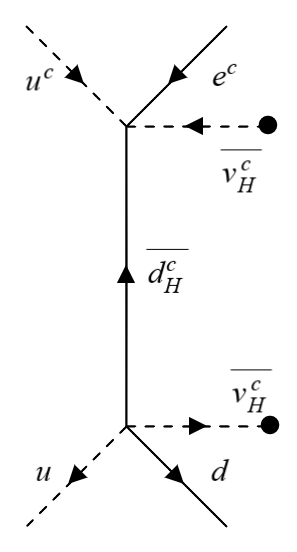}} \\
\subfloat[]{\includegraphics[scale=0.32]{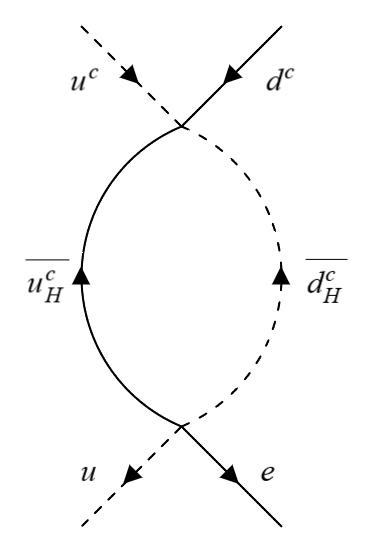}} \hspace{0.5cm} &\subfloat[]{\includegraphics[scale=0.32]{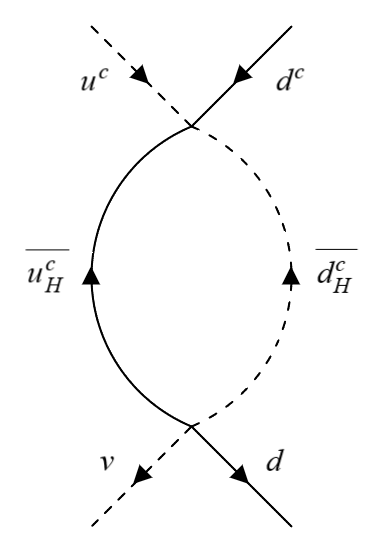}} \hspace{0.5cm}\hspace{0.5cm}
\end{tabular} 
\end{tabularx}
\caption{\label{pd5bar} Dimension-five proton decay diagrams. Dashed lines represent bosons, solid lines represent fermions, and dots represent the vevs. These are graphs between the $B$-violating superpotential coupling $FF\overline{H^c}\ \overline{H^c}$ and the $B$-conserving superpotential coupling $F^cF^c\overline{H^c}\ \overline{H^c}$.}
\end{figure}

\begin{figure}[th]
\begin{tabularx}{0.8\textwidth}{@{}cXX@{}}
\begin{tabular}{ccc}
\subfloat[]{\includegraphics[scale=0.32]{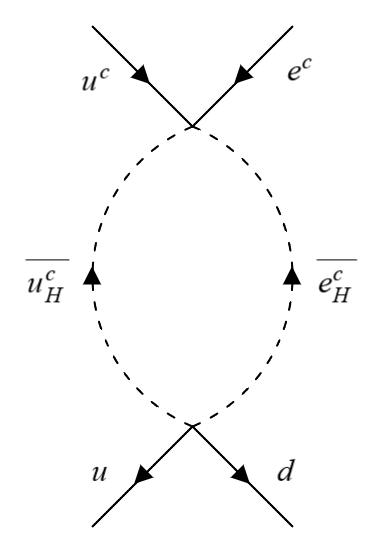}} \hspace{0.5cm}&\subfloat[]{\includegraphics[scale=0.32]{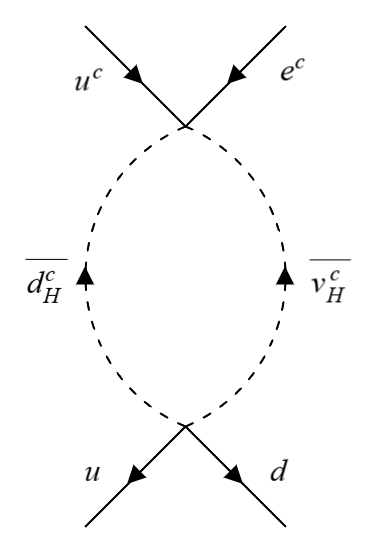}}  \hspace{0.5cm} & \subfloat[]{\includegraphics[scale=0.52]{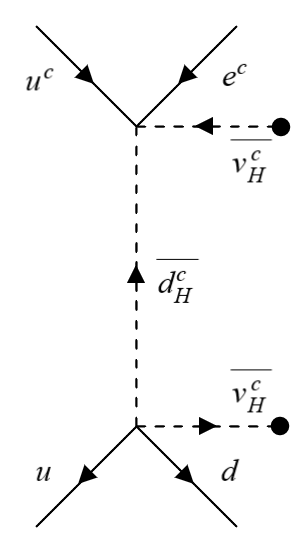}} \\
\subfloat[]{\includegraphics[scale=0.32]{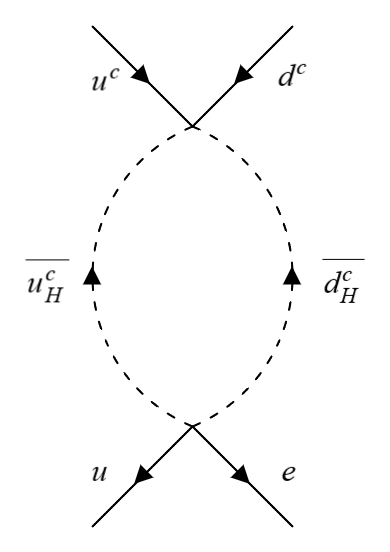}} \hspace{0.5cm} &\subfloat[]{\includegraphics[scale=0.32]{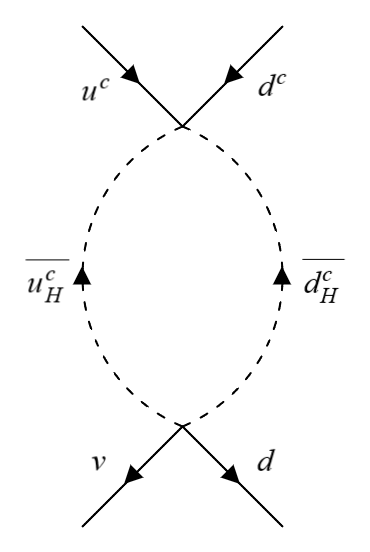}}\hspace{0.5cm}\hspace{0.5cm}   
\end{tabular}
\end{tabularx}
\caption{\label{pd6bar} Dimension-six proton decay diagrams. Dashed lines represent bosons, solid lines represent fermions, and dots represent the vevs. These are graphs between the 
$B$-violating superpotential coupling $FF\overline{H^c}\ \overline{H^c}$ and the $B$-conserving superpotential coupling  
$F^cF^c\overline{H^c}\ \overline{H^c}$.}
\end{figure}

\begin{figure}[htbp!]
\begin{tabularx}{0.8\textwidth}{@{}cXX@{}}
\begin{tabular}{ccc}
\subfloat[]{\includegraphics[scale=0.32]{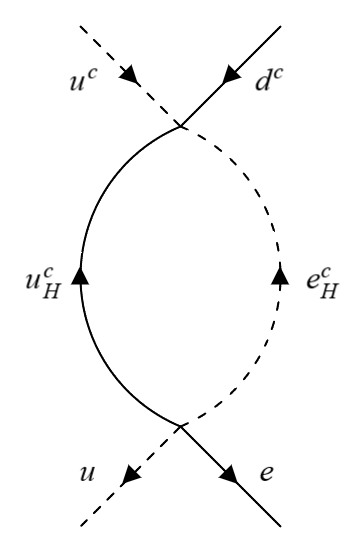}}\hspace{0.5cm} &\subfloat[]{\includegraphics[scale=0.32]{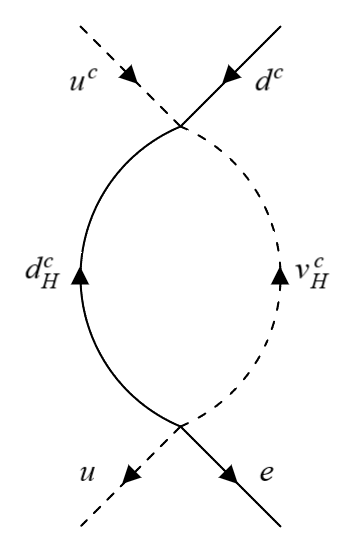}} \hspace{0.5cm} & \subfloat[]{\includegraphics[scale=0.52]{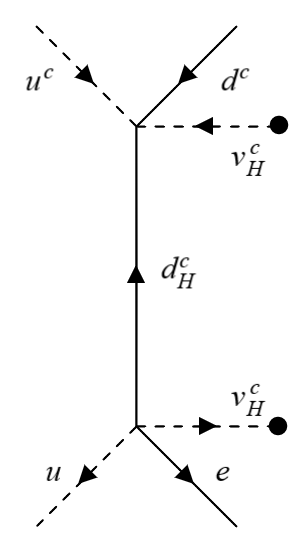}} \\
\subfloat[]{\includegraphics[scale=0.32]{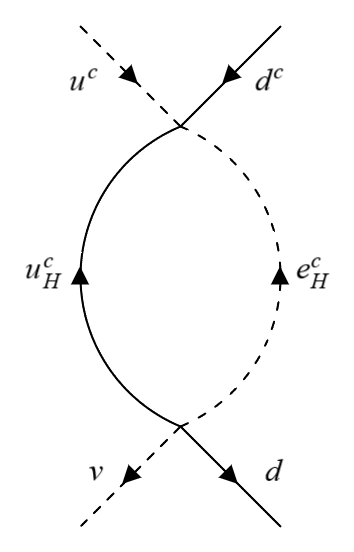}} \hspace{0.5cm}&\subfloat[]{\includegraphics[scale=0.32]{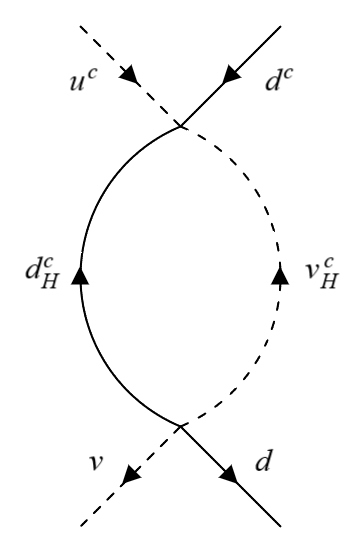}}\hspace{0.5cm} & \subfloat[]{\includegraphics[scale=0.52]{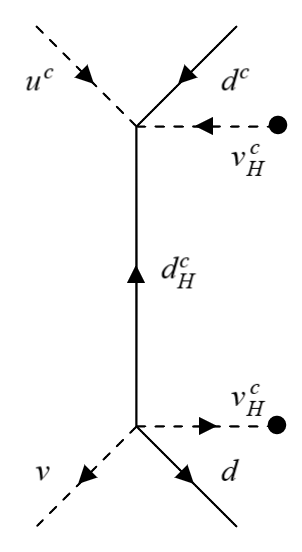}}  \\
&\subfloat[]{\includegraphics[scale=0.32]{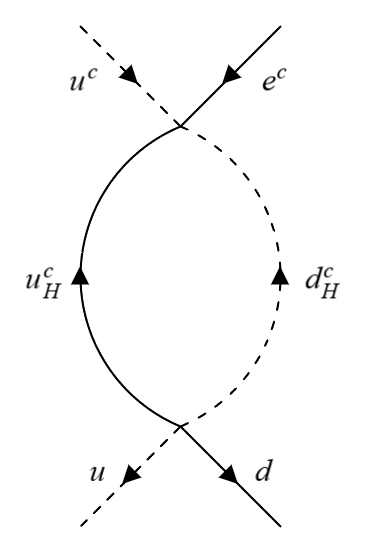}}& 
\end{tabular}
\end{tabularx}
\caption{\label{pd5} Dimension-five proton decay diagrams. Dashed lines represent bosons, solid lines represent fermions, and dots represent the vevs. These are graphs between the $B$-violating superpotential coupling $F^cF^cH^cH^c$ and the $B$-conserving superpotential coupling  
$FFH^cH^c$.}
\end{figure}

\begin{figure}[htbp!]
\begin{tabularx}{0.8\textwidth}{@{}cXX@{}}
\begin{tabular}{ccc}
\subfloat[]{\includegraphics[scale=0.32]{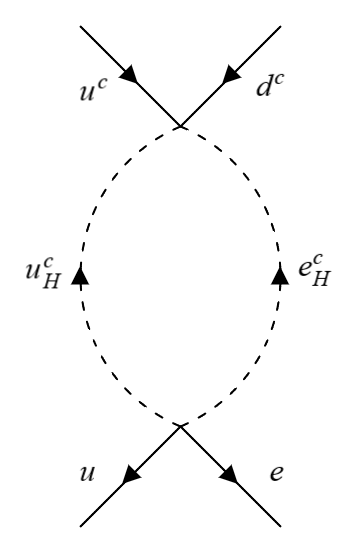}} \hspace{0.5cm}&\subfloat[]{\includegraphics[scale=0.32]{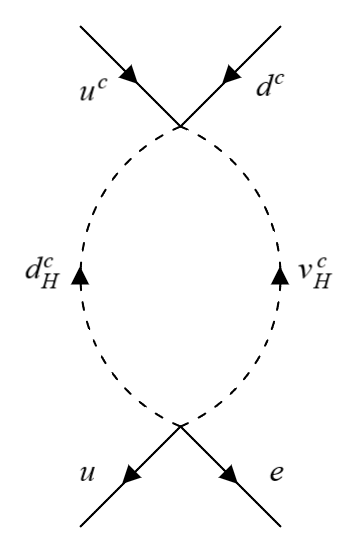}}  \hspace{0.5cm} & \subfloat[]{\includegraphics[scale=0.52]{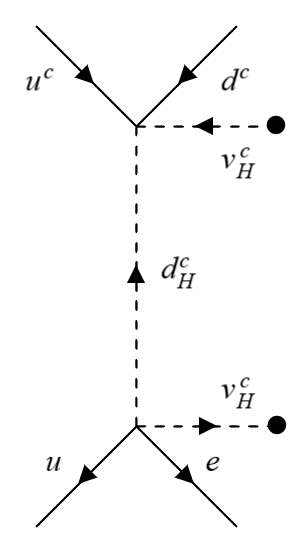}} \\
\subfloat[]{\includegraphics[scale=0.32]{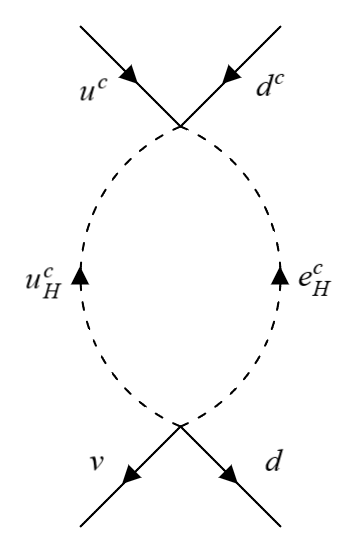}} \hspace{0.5cm} &\subfloat[]{\includegraphics[scale=0.32]{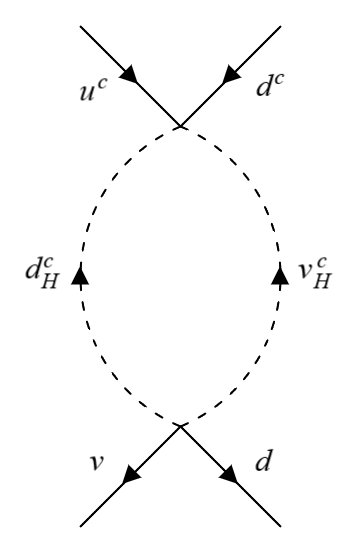}}\hspace{0.5cm}& \subfloat[]{\includegraphics[scale=0.52]{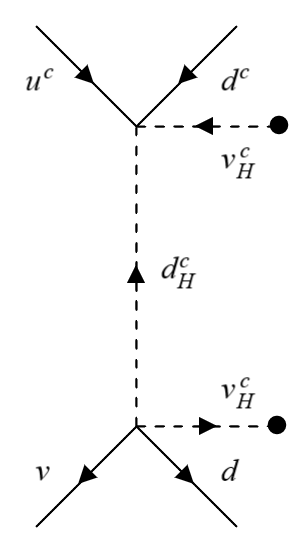}} \\
  & \subfloat[]{\includegraphics[scale=0.32]{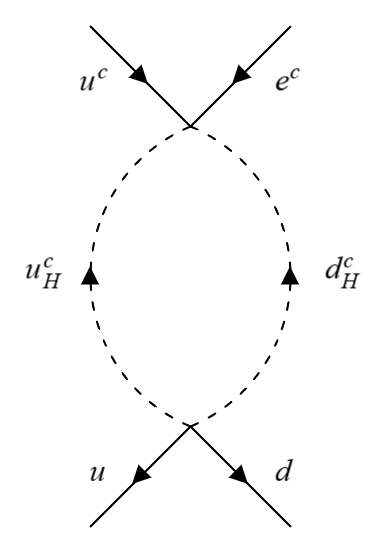}}\hspace{0.5cm}&   
\end{tabular}
\end{tabularx}
\caption{\label{pd6} Dimension-six proton decay diagrams. Dashed lines represent bosons, solid lines represent fermions, and dots represent the vevs. These are graphs between the 
$B$-violating superpotential coupling $F^cF^cH^cH^c$ and the $B$-conserving superpotential coupling  
$FFH^cH^c$.}
\end{figure}

Once again the couplings $F^cF^cH^cH^c$, $FFH^cH^c$ and $F^cF^c\overline {H^c}\ \overline {H^c}$, $FF\overline{H^c}\ \overline{H^c}$ defined in Eq.~(\ref{gij}) play crucial role for the realization of proton decay corresponding to the operators described in Eq.~(\ref{LLRR}). The Feynman diagrams for dimension-five and dimension-six operators corresponding to the couplings $FF\overline{H^c}\ \overline{H^c}$, $F^cF^c\overline{H^c}\ \overline{H^c}$ are shown in Fig.~\ref{pd5bar} and Fig.~\ref{pd6bar}, respectively. Analogous diagrams for the couplings $F^cF^cH^cH^c$, $FFH^cH^c$ are shown in Fig.~\ref{pd5} and Fig.~\ref{pd6}. 
It is important to notice that the internal fermion lines represent the chirality nonchanging part of the fermion propagator $\slashed{p}/(p^2-m^2)$. For dimension-five proton decay diagrams (Figs.~\ref{pd5bar} and \ref{pd5}), the fermionic or bosonic character of the external lines in each vertex can be interchanged independently. Also the fermionic or bosonic character of the lines in the
loops can be interchanged independently.

The loop diagrams in Figs.~\ref{pd5bar}, \ref{pd6bar}, \ref{pd5}, and \ref{pd6} are expected to make somewhat smaller contribution than the tree ones because of the loop factors. Therefore, we will only concentrate on the tree diagrams. For proton decay via dimension-five diagrams, we must form a loop by connecting the two external bosons by a Higgsino or gaugino line to turn them into external fermions. This line will not involve chirality flipping and thus will be of the type $\slashed{p}/(p^2-m^2)$. So the loop integral will be $\sim 1/m_{d_H^c}^2$ or $\sim 1/m_{\overline{d_H^c}}^2\,$, as the case may be, multiplied by logarithms and loop factors. Therefore, its contribution is relatively suppressed as compared to the contribution of the conventional dimension-five proton decay diagram with chirality flipping color-triplet Higgs exchange. Note that the dimension-six tree diagrams in Figs.~\ref{pd6bar} and \ref{pd6} come without the logarithms and the loop factors, and so their contribution is expected to be dominant unless the logarithms are very significant. Therefore, we only focus on the dimension-six tree diagrams of Figs.~\ref{pd6bar} and \ref{pd6} with the following decay rates:
\begin{align}
\Gamma (p\rightarrow \pi^0 l^+_i) &\simeq C_{\pi}
\left( \frac{v}{\Lambda}\right)^4 \left| A_{\pi l^+_i}\right|^2 \left( \left|  \frac{(\gamma_1)^{\dagger}_{11}(\gamma_2)_{1i}}{m_{d_H^c}^2}\right|^2 +\left| \frac{(\overline{\gamma}_{2})_{11}(\overline{\gamma}_{1})^{\dagger}_{1i}}{m_{\overline{d_H^c}}^2}\right|^2 \right), \label{pil} \\
\Gamma (p\rightarrow \pi^+ \overline{\nu}_i) &\simeq C_{\pi} \left( \frac{v}{\Lambda}\right)^4 \left| A_{\pi \overline{\nu}_i}\right|^2
 \left|  \frac{(\gamma_1)^{\dagger}_{11}(\gamma_2)_{1i}}{m_{d_H^c}^2}\right|^2 , \label{pinu}\\
\Gamma (p\rightarrow K^0 l^+_i) &\simeq C_K
\left( \frac{v}{\Lambda}\right)^4 \left| A_{K l^+_i}\right|^2 \left( \left|  \frac{(\gamma_1)^{\dagger}_{12}(\gamma_2)_{1i}}{m_{d_H^c}^2}\right|^2 +\left| \frac{(\overline{\gamma}_{2})_{12}(\overline{\gamma}_{1})^{\dagger}_{1i}}{m_{\overline{d_H^c}}^2}\right|^2 \right), \label{Kl} 
\end{align}
\begin{align}
\Gamma (p\rightarrow K^+ \overline{\nu}_i) &\simeq C_K \left( \frac{v}{\Lambda}\right)^4 
\left( \left| A^R_{K \overline{\nu}_i}\right|^2 \left|  \frac{(\gamma_1)^{\dagger}_{12}(\gamma_2)_{1i}}{m_{d_H^c}^2} \right|^2 + \left| A^L_{K \overline{\nu}_i}\right|^2 \left| \frac{(\gamma_1)^{\dagger}_{11}(\gamma_2)_{2i}}{m_{d_H^c}^2}\right|^2 \right), \label{Knu}
\end{align}
with
\begin{equation}
C_{\pi} = \frac{m_p}{32\pi} \left(1 - \frac{m_{\pi}^2}{m_p^2} \right)^2, \quad  C_K = \frac{m_p}{32\pi} \left(1 - \frac{m_{K}^2}{m_p^2} \right)^2.
\end{equation}
Here, $m_p$, $m_{\pi}$, and $m_{K}$ are the proton, pion, and kaon mass respectively, and $l^+_i=e^+\,{\rm or}\, \mu^+$. 

The hadronic matrix elements $(A_{\pi e^+},\,A_{\pi \mu^+}) = (-0.131,\,-0.118)$ GeV$^2$, $A_{\pi \overline{\nu}_i} = -0.186$ GeV$^2$, $(A_{K e^+},\,A_{K \mu^+}) = (-0.103,\,-0.099)$ GeV$^2$, and $(A^R_{K \overline{\nu}_i},\,A^L_{K \overline{\nu}_i}) = (-0.049,\,-0.134)$ GeV$^2$ are assigned their recently updated values from lattice computations \cite{Aoki:2017puj}. It is interesting to note that the value of $a$ ($b$) or the mass of $d_H^c$ ($\overline{d_H^c}$) can be made small enough to reduce the proton lifetime to a measurable level. The current limits on the proton lifetime for the various decay modes mentioned above are $\tau (p\rightarrow \pi^0 (e^+,\,\mu^+)) > (16,\,7.7) \times 10^{33}$ yrs \cite{Miura:2016krn},
$\tau (p\rightarrow \pi^+ \overline{\nu}) > 3.9 \times 10^{32}$ yrs \cite{Abe:2013lua}, $\tau (p\rightarrow K^0 (e^+,\,\mu^+)) > (1,\,1.6) \times 10^{33}$ yrs \cite{Kobayashi:2005pe,Regis:2012sn}, and $\tau (p\rightarrow K^+ \overline{\nu}) > 6.6 \times 10^{33}$ yrs \cite{Abe:2014mwa,Takhistov:2016eqm}. With all $\gamma_{1,2}^{ij}$'s and $\overline{\gamma}_{1,2}^{ij}$'s in Eq.~(\ref{gij}) being of order $\gamma_i$, the decay mode $p\rightarrow e^+ \pi^0$ provides the following most stringent bound on the masses of the color triplets:
\begin{equation} \label{ab2}
\frac{m_{d_H^c}}{M_i} \text{ and/or } \frac{m_{\overline{d_H^c}}}{M_i} \gtrsim  0.17 \left(\frac{2\times 10^{16} \text{ GeV}}{v} \right).
\end{equation}
Therefore, with $M_i=10^{14}$ GeV and $v = 2 \times 10^{16}$ GeV, we obtain $m_{d_H^c} \text{ and/or } m_{\overline{d_H^c}} \gtrsim 1.7 \times 10^{13}$ GeV or $a \text{ and/or } b \gtrsim 8\times 10^{-4}$. Thus the bound obtained from the chirality nonflipping class of dimension-six operators is far more stringent  compared to the one obtained earlier from the chirality flipping dimension-five and dimension-six diagrams. Assuming natural values for the couplings $a,\,b \sim 1$, the corresponding decay rate becomes comparable to the one from the dimension-six operator $F F (F^c)^{\dagger}(F^c)^{\dagger}/\Lambda^2$ obtained from the same nonrenormalizable term in the K\"ahler potential. However, it is relatively suppressed compared to the gauge boson mediated dimension-six proton decay rate in a typical GUT model.

It is instructive to estimate a few important branching fractions in order to make a comparison of the present model with the other GUT models. To do this, we assume all $\gamma_{1,2}^{ij}$'s and $\overline{\gamma}_{1,2}^{ij}$'s in Eq.~(\ref{gij}) to be of order $\gamma_i$ with $M_i \sim \gamma_i (v^2/\Lambda)$. Using Eqs.~(\ref{pil})-(\ref{Knu}), the following relevant branching fractions can be obtained:
\begin{eqnarray}
\frac{\Gamma (p\rightarrow \pi^0 \mu^+)}{\Gamma (p\rightarrow \pi^0 e^+)} &\simeq& 0.81, \quad \frac{\Gamma (p\rightarrow K^0 e^+)}{\Gamma (p\rightarrow \pi^0 e^+)} \simeq 0.34, \quad \frac{\Gamma (p\rightarrow K^0 \mu^+)}{\Gamma (p\rightarrow \pi^0 \mu^+)} \simeq 0.39,   \\
\frac{\sum_i\Gamma (p\rightarrow \pi^+ \overline{\nu}_i)}{\Gamma (p\rightarrow \pi^0 e^+)} &\simeq&  \frac{6.06}{1 + \frac{m_{d_H^c}^2}{m_{\overline{d_H^c}}^2}}, \qquad \frac{\sum_i\Gamma (p\rightarrow \pi^+ \overline{\nu}_i)}{\sum_i\Gamma (p\rightarrow K^+ \overline{\nu}_i)} \simeq  3.11.
\end{eqnarray}
These predictions can be compared, for example, with the predictions of the no-scale supersymmetric standard unflipped S$U(5)$ and flipped $SU(5)$ models recently calculated in Ref.~\cite{Ellis:2020qad}.
Also see Refs.~\cite{Babu:1997js,Babu:1998wi,Haba:2020bls} for $SO(10)$ models. Most of the above  branching fractions lie, in our case, close to unity except for $\sum_i\Gamma (p\rightarrow \pi^+ \overline{\nu}_i)/ \Gamma (p\rightarrow \pi^0 e^+)$, which lies between 6.06 and  6.06$(m_{\overline{d_H^c}}/ m_{d_H^c})^2$ for  $m_{d_H^c} \ll m_{\overline{d_H^c}}$ and $m_{d_H^c} \gg m_{\overline{d_H^c}}$, respectively.
These are very distinctive predictions which are expected to be tested in future experiments.

Most of the previous work on the important topic of proton decay in the $G_{4-2-2}$ model is based on the nonsupersymmetric version of this model. For example, in Refs.~\cite{Pati:1983zp,Pati:1983jk,Saad:2017pqj}, proton decay is discussed by employing the `minimal' Higgs content with the $(15,2,2)$ Higgs multiplet playing a crucial role in realizing the important proton decay modes. As pointed out in Ref.~\cite{Mohapatra:1980qe}, a `minimal' Higgs content without the $(15,2,2)$ multiplet does not lead to proton decay. It is shown that the $(B-L)$-nonconserving dimension-nine and dimension-ten operators require a symmetry breaking scale of order 100~TeV or lower for proton decay to be in the observable range. This is in contrast to the present model, which is mostly based on the Higgs content employed in the original $G_{4-2-2}$ model \cite{Pati:1974yy}. Here the $(B-L)$-conserving dimension-five and dimension-six operators lead to observable proton decay modes with the $G_{4-2-2}$ symmetry breaking scale of order $M_{GUT}$.

\section{Gauge Coupling Unification}

It is important to emphasize that with $a \text{ or }\,b\sim 10^{-3}$ the proton lifetime is predicted within the potentially observable range of Hyper-Kamiokande, $\tau (p\rightarrow e^+ \pi^0) < 7.8 \times 10^{34}$ yrs \cite{Abe:2018uyc}. The corresponding values of the color-triplet masses $m_{(d_H^c,\,g)} \text{ or } \, m_{(\overline{d_H^c}, \,g^c)}$ are $\sim 10^{13}$~GeV, and therefore lie somewhat below the GUT scale.
These reduced masses ultimately ruin gauge coupling unification, an attractive feature of MSSM.
As $G_{\text{4-2-2}}$ is a semi-simple group, gauge unification is not a must, but it can, in any case, be achieved with a modest adjustment of the model. As an example, let us consider the two color-triplet fields $d_H^c,\,g$ and $\overline{d_H^c},\,g^c$ to be both of intermediate mass $\sim 10^{13}$~GeV with $a \text{ and }b\sim 10^{-3}$. To recover gauge coupling unification we add an arbitrary number of bi-doublets $\mathfrak{H}_{\alpha}=\mathfrak{H}_1, \, \mathfrak{H}_2, \, \mathfrak{H}_2,\cdots$, with $R$ charge $R(\mathfrak{H}_{\alpha})=1$. To avoid any unnecessary couplings of these bi-boublets with the MSSM matter superfields $F$, $F^c$, we further assume an additional discrete $Z_2$ symmetry under which only the $\mathfrak{H}_{\alpha}$'s are odd, and all the other superfields are even. This symmetry remains unbroken and thus does not lead to a domain wall problem. 

\begin{figure}[htbp!]
\begin{tabularx}{2.0\textwidth}{@{}cXX@{}}
\begin{tabular}{cc}
\subfloat[]{\includegraphics[scale=0.65]{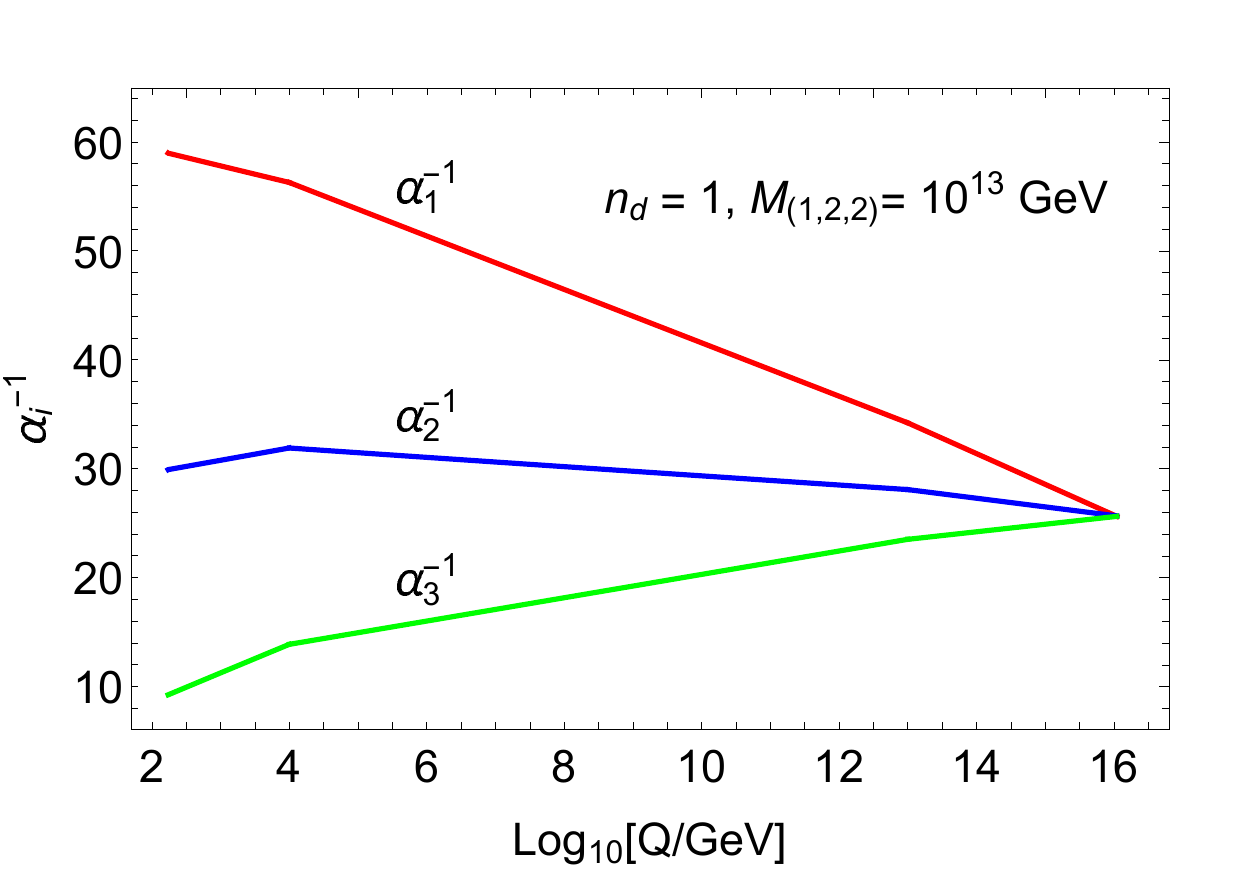}}&\subfloat[]{\includegraphics[scale=0.65]{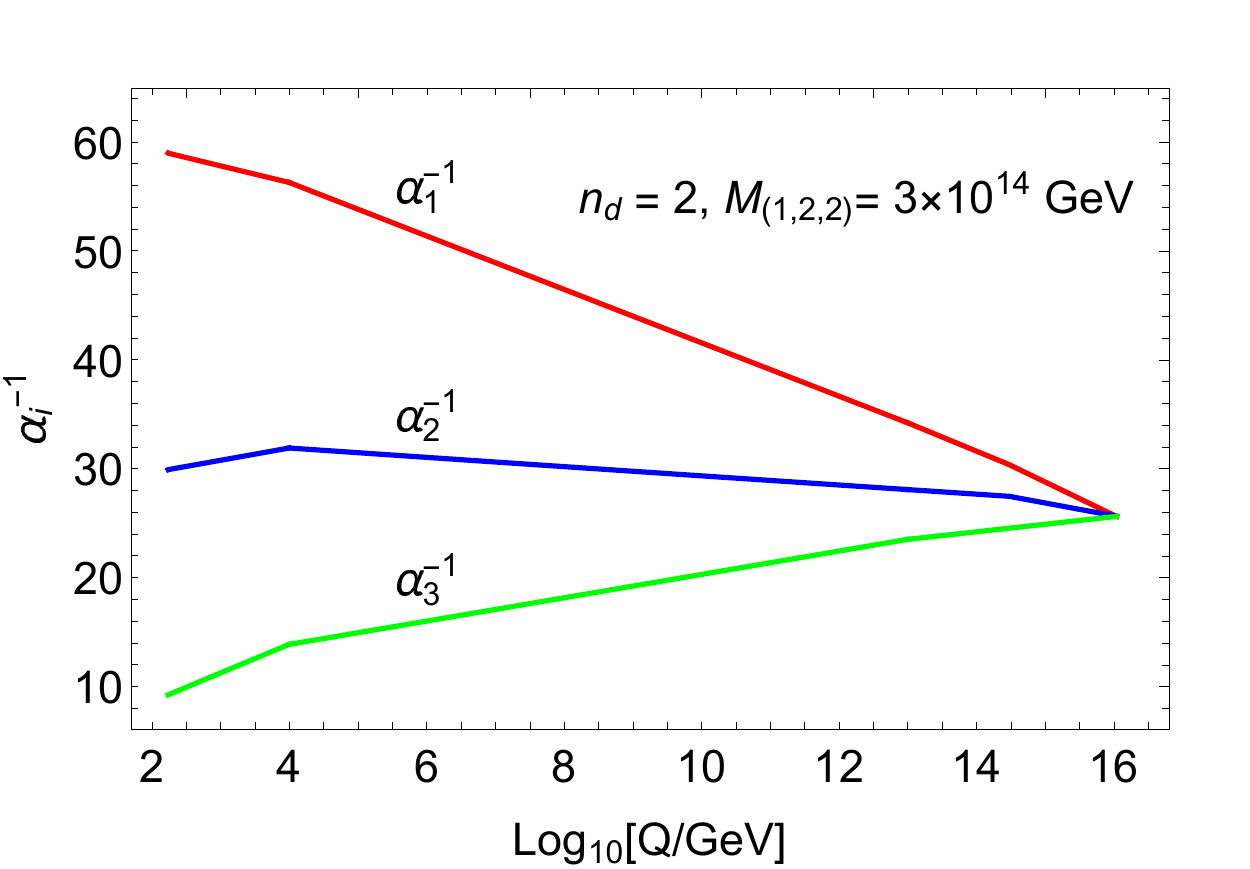}}  \\
\subfloat[]{\includegraphics[scale=0.65]{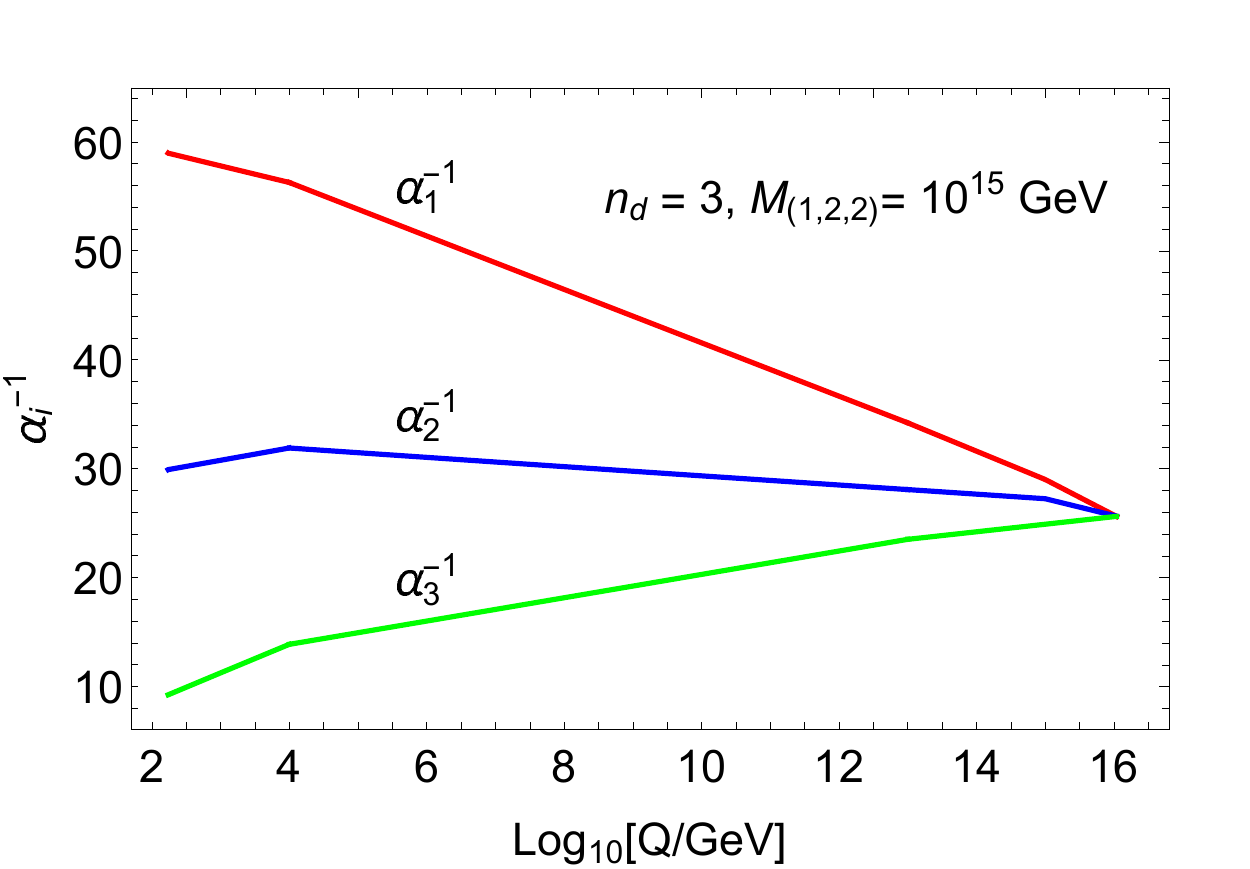}} &\subfloat[]{\includegraphics[scale=0.65]{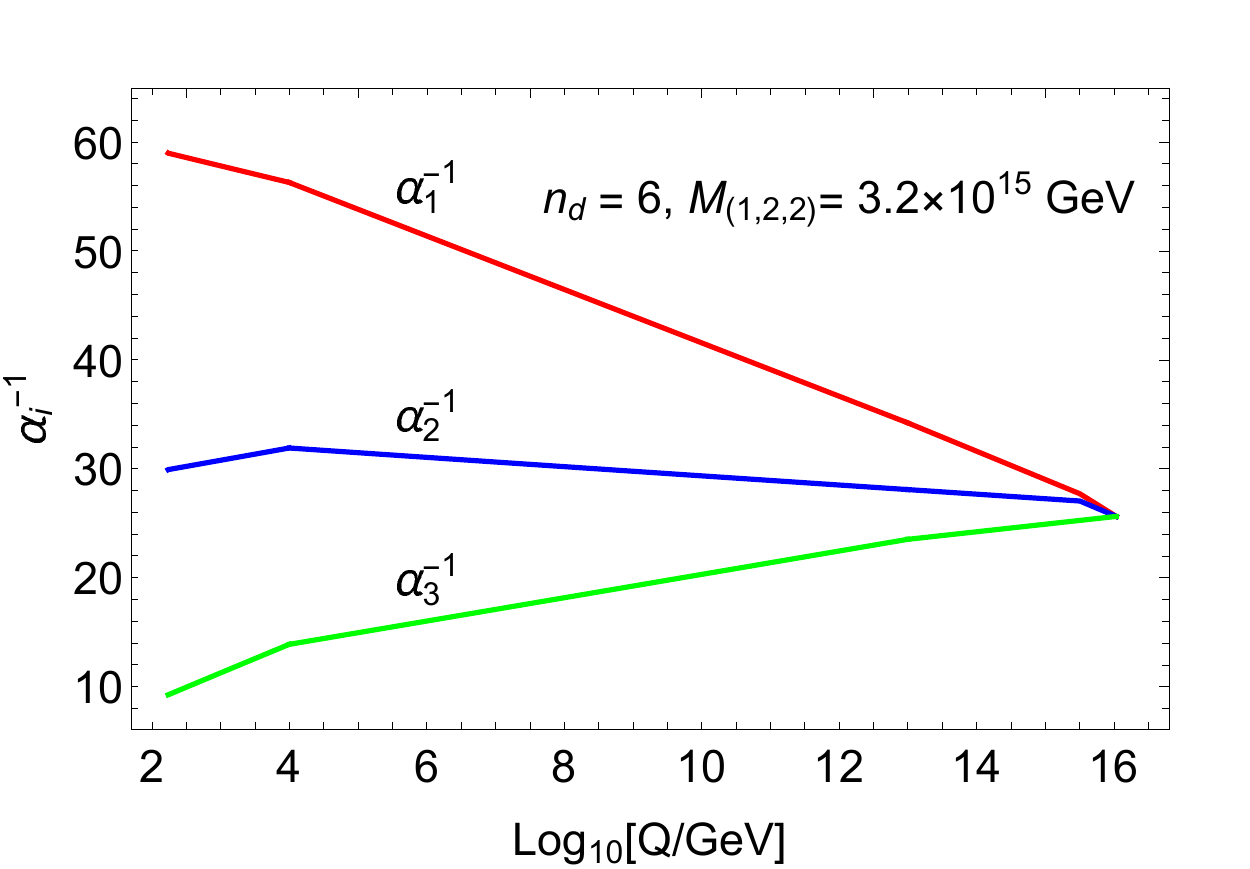}} 
\end{tabular}
\end{tabularx}
\caption{\label{gcu} The evolution of the inverse gauge couplings $\alpha^{-1}_i = 4\pi/g_i^2$ versus the energy scale $Q$ in the $R$-symmetric $G_{\text{4-2-2}}$ model where the two color-triplet fields $d_H^c,\,g$ and $\overline{d_H^c},\,g^c$ are taken to be of intermediate mass $\sim 10^{13}$~GeV, appropriate for potentially observable proton decay. The gauge-coupling unification is shown for four choices with $n_d=1,2,3,6$ number of additional bi-doublets of mass $M_{(1,2,2)}=10^{13},\,3\times 10^{14},\,10^{15},\,3.2 \times 10^{15}$~GeV, respectively.}
\end{figure}

The general form of the allowed nonrenormalizable superpotential terms involving the $\mathfrak{H}_{\alpha}$ superfields is
\begin{equation}
\epsilon \Lambda \, \mathfrak{H}^2
\left(\frac{h^2}{\Lambda^2} \right)^{m} \left(\frac{\overline{H^c}H^c}{\Lambda^2}  \right)^n
\left(\frac{(\overline{H^c})^4}{\Lambda^4}  \right)^{p}
\left(\frac{(H^c)^4}{\Lambda^4}  \right)^{q},
\end{equation}
with $m,n,p,q = 0,1,2,\cdots$, and with at least one of them being nonzero. Here, the indices on $\mathfrak{H}^2$ and the dimensionless constant $\epsilon$ are suppressed. On the other hand, the only allowed renormalizable terms involvings these extra bi-doublets are their mass terms $M_{\alpha\beta}\mathfrak{H}_{\alpha}\mathfrak{H}_{\beta}$, which can be chosen at will. The leading nonrenormalizable terms $\mathfrak{H}^2 H^c\overline{H^c}/\Lambda$ can provide additional intermediate scale contributions to the masses of the extra bi-doublets. The overall masses of these fields can then be appropriately adjusted so as to achieve successful gauge coupling unification. Four choices for the number of the additional bi-doublets $n_d=1,\,2,\,3,\,6$ are shown in Fig.~\ref{gcu}, where a successful gauge coupling unification is achieved if their common mass is $M_{(1,2,2)}=10^{13},\,3 \times 10^{14},\,10^{15},\,3.2\times 10^{15}$~GeV, respectively. Therefore, with a suitable choice of $n_d$ we can obtain $M_{(1,2,2)}$ values close to the GUT scale.  Finally, the $Z_2$ symmetry also makes the additional bi-doublets stable (as they can only annihilate in pairs) and thus provides potential candidates for dark matter of intermediate mass scale. For a recent discussion of intermediate mass fermionic dark matter see \cite{Lazarides:2020frf}.

\section{\label{con}Conclusion}
We have considered proton decay in a class of realistic supersymmetric $SU(4)_c \times SU(2)_L \times SU(2)_R$ models. The basic structure of the model is determined by implementing supersymmetric hybrid inflation such that the monopole problem is adequately resolved, the low energy sector coincides with the MSSM, and the neutrinos have the desired masses to explain the observed neutrino oscillations. Proton decay is mediated by color triplets present in the various chiral superfields, and it lies within the reach of detectors such as Hyper-Kamiokande for a range of intermediate scale masses of these color triplets. Unification of the MSSM gauge couplings in the presence of such color triplets is an important issue which is also discussed.
\section*{Acknowledgments}
The work of G.L. and Q.S. was supported by the Hellenic Foundation for Research
and Innovation (H.F.R.I.) under the ``First Call for H.F.R.I. Research Projects to support Faculty Members and Researchers and the procurement of high-cost research equipment grant'' (Project 
Number:2251). This work is partially supported by the DOE grant No. DE-SC0013880 (Q.S.). We thank Fariha Vardag and Maria Mehmood for their help with the figures.

\end{document}